\documentclass{PoS}

\usepackage{amsmath,bbm}
\usepackage{tikz}

\newcommand{\id}{\mathbbm{1}}

\newcommand{\U}{\mathrm{U}}
\newcommand{\Tr}{\operatorname{Tr}}

\newcommand{\sgn}{\operatorname{sgn}}
\newcommand{\img}{\operatorname{Im}}
\newcommand{\real}{\operatorname{Re}}

\title{Domain-wall, overlap, and topological insulators}

\ShortTitle{Domain-wall, overlap, and topological insulators}

\author{\speaker{Taro Kimura}
        \\
        Department of Physics, Keio University \\
	        E-mail: \email{taro.kimura@keio.jp}
	}

\abstract{
Topological insulators are a new class of materials which have gapped
spectra in the bulk, but are accompanied by topologically protected
gapless excitations at the surface (edge) of the system. 
These phenomena have a close relationship with symmetry and
dimensionality of the system through quantum anomalies. 
We point out that such a surface state is a physical realization of the
domain-wall/overlap fermion. 
From this point of view, we discuss its implications for experiments of
topological insulators. 
We also discuss an unconventional overlap fermion, which is suggested by
the ``periodic table'' of topological insulators and superconductors.
}

\FullConference{
         The 33rd International Symposium on Lattice Field Theory\\
	 14-18 July 2015\\
	 Kobe International Conference Center, Kobe, Japan}

\begin{document}

\section{Introduction}

Construction of the doubler-free chiral fermion on a lattice has been a longstanding problem in lattice field theory.
Nowadays it is known that the domain-wall fermion~\cite{Kaplan1992} and the overlap fermion~\cite{Neuberger1998} present a solution to this problem, and a lot of theoretical and numerical works are performed with these lattice fermions.
In this report, we discuss a potential application of these formalisms to condensed-matter physics, especially topological insulators, where topology and quantum anomaly play an important role in the characterization of the system.
We especially show that the topological insulator edge (surface) state is a physical realization of the domain-wall/overlap fermion.

\section{Band topology and mass gap}
\label{sec:band}


Let us explain how the mass term plays a role in the band topology.
We start with the two-dimensional massive Dirac Hamiltonian, which is a toy model for the two-dimensional quantum Hall effect, classified into the class A system~\cite{Schnyder2008,Kitaev2009},
\begin{align}
 H_\text{2D}(p)
 = p_x \sigma_x + p_y \sigma_y + m \sigma_z
 =
 \left(
  \begin{array}{cc}
   m & \Delta(p)^\dag \\ \Delta(p) & -m
  \end{array}
 \right)
 \label{eq:2D_Ham}
\end{align}
where $\Delta(p) = p_x + i p_y \in \mathbb{C}$.
This Hamiltonian is actually given as the Bloch Hamiltonian of the real-space Hamiltonian.
This is also seen as a Hermitian version of the Dirac operator through multiplication of $\sigma_z$-matrix, $\displaystyle H_\text{2D} = \sigma_z \, D_\text{2D}$.

The Hamiltonian \eqref{eq:2D_Ham} has two eigenvalues
\begin{align}
 \lambda^{(\pm)} = \pm \sqrt{p_x^2 + p_y^2 + m^2}
 \, ,
\end{align}
and we can define the topological charge associated with each eigenstate.
For the positive eigenvalue state, it is given by
\begin{align}
 \nu_\text{2D} = \frac{i}{2\pi}
 \int dp_x dp_y \ F
 = \frac{1}{2} \operatorname{sgn} (m)
 \, ,
 \label{eq:2D_top}
\end{align}
while an extra factor $(-1)$ is needed for the negative eigenvalue state.
This is called the TKNN number, 
which computes the Hall conductivity of the model.
Now the integral \eqref{eq:2D_top} provides the first Chern number, which is topological, such that its value depends only on the sign of the mass parameter.
The curvature $F$, called the Berry curvature, is defined with the Berry connection using the eigenfunction of the Hamiltonian~\eqref{eq:2D_Ham}.
Let us show the explicit expressions for latter convenience,
\begin{align}
  \psi
 & =
 \frac{1}{\sqrt{1 + |\xi|^2}}
 \left( \begin{array}{c} \xi \\ 1 \end{array} \right) \, ,
 \qquad
 A = \psi^\dag d \psi
 = i \img \frac{\xi^\dag d \xi}{1 + |\xi|^2} \, ,
 \qquad
 F = \frac{d\xi^\dag d\xi}{\left( 1 + |\xi|^2 \right)^2} \, ,
 \label{eq:forms}
\end{align}
where we apply the differential form notation to the connection and the curvature, and the derivative is with respect to momentum, $d = \partial_{p_i} dp^i$.
The complex parameter $\xi$ is now given by
\begin{align}
 \xi
 = \frac{\lambda^{(\pm)} + m}{\Delta}
 = \frac{\Delta^\dag}{\lambda^{(\pm)} - m}
 \, .
 \label{eq:CP1_coord}
\end{align}
This system enjoys the $\U(1)$ gauge symmetry, corresponding to the phase rotation of $\xi$.
This $\U(1)$ symmetry reflects the particle number conservation for each energy band, which is well-defined as long as the mass parameter takes a non-zero value $m \neq 0$.
This is nothing but a consequence of the adiabatic approximation in the original sense of the Berry phase.

Let us comment on the reason why the topological charge \eqref{eq:2D_top} takes a half-integer value, while its difference must be integer $\delta \nu_\text{2D} = \pm 1$, which occurs at the massless point $m=0$~\cite{Oshikawa1994}.
Usually the integral like \eqref{eq:2D_top} is evaluated with the asymptotic behavior of the curvature and the connection at infinity $p \to \infty$, which plays a role of the boundary.
In this case, however, the origin of the momentum space $p = 0$ also plays a similar role, 
because the parameter $\xi$ takes a non-trivial value at $p=0$.
This also reflects the anomaly of the (2+1)-dimensional Dirac fermion, namely the parity anomaly.
We also remark that the construction of the topological charge using the fermion, or more precisely, the Clifford algebra, e.g. Pauli matrices, is known as the Atiyah--Bott--Shapiro construction~\cite{Atiyah1964}, which is essentially related to the K-theoretical point of view.


The argument above is straightforwardly generalized to the four-dimensional case with the following Hamiltonian, which again belongs to the class A system (the four-dimensional quantum Hall system),
\begin{align}
 H_\text{4D}(p)
 = p \cdot \gamma + m \gamma_5
 =
 \left(
  \begin{array}{cc}
   m & \Delta(p)^\dag \\ \Delta(p) & -m
  \end{array}
 \right) 
 \label{eq:4D_Ham}
\end{align}
where each matrix element is 2-by-2, so that the Hamiltonian itself is a 4-by-4 matrix.
This is again related to the Dirac operator, $\displaystyle H_\text{4D} = \gamma_5 \, D_\text{4D}$.
The off-diagonal element is given by $\Delta(p) = p \cdot \sigma$ with the four-vector $\sigma = (\id,\vec{\sigma})$, which takes a quaternionic value $\Delta(p) \in \mathbb{H}$.
We can formally apply the same expression as in two dimensions \eqref{eq:forms} to the four-dimensional model \eqref{eq:4D_Ham}.
This Hamiltonian has two eigenvalues
\begin{align}
 \lambda^{(\pm)}
 & =
 \pm \sqrt{p^2 + m^2}
 \, ,
\end{align}
and each eigenstate is degenerated twice.
Thus the Berry connection becomes a 2-by-2 matrix, and the corresponding topological charge is now given by
\begin{align}
 \nu_\text{4D}
 & =
 - \frac{1}{16\pi^2} \int d^4 p \, \Tr F * F
 = \frac{1}{2} \sgn(m)
 \, .
 \label{eq:4D_top}
\end{align}
This implies that the Berry connection realizes the four-dimensional instanton configuration in the momentum space.%
\footnote{
This construction can be actually discussed in parallel with the systematic construction of instantons, called the ADHM construction~\cite{Hashimoto2015}.
}
The topological number \eqref{eq:4D_top} takes a half-integer value, due to the same reason as the two-dimensional case, and its difference is again given by $\delta \nu_\text{4D} = \pm 1$ at $m=0$.


We can discuss the topological charge also with the lattice system.
We define the Hermitian operator, associated with the four-dimensional Wilson fermion
\begin{align}
 H_\text{W}(p) = \gamma_5 \, D_\text{W}(p)
 \, , \qquad
 D_\text{W}(p)
 & =
 m
 + \sum_{\mu=1}^4 i \gamma_\mu \sin p_\mu
 + r \sum_{\mu=1}^4 (1 - \cos p_\mu)
 \, .
\end{align}
Since it is a lattice system, the topological charge is then defined as an integral over the Brillouin zone~\cite{Qi2008},
\begin{align}
 \nu_\text{4D}
 & =
 - \frac{1}{16\pi^2} \int_\text{BZ} d^4 p \, \Tr F * F
 =
 \begin{cases}
  0 & (0 < m \ \& \ m < -8r) \\
  1 & (-2r < m < 0) \\
  -3 & (-4r < m < -2r) \\
  3 & (-6r < m < -4r) \\
  -1 & (-8r < m < -6r) \\  
 \end{cases}
 \, .
 \label{eq:4D_top_lat}
\end{align}
In this case, the topological number itself takes an integral value, and its change $\delta \nu_\text{4D} = +1$, $-4$, $+6$, $-4$, $+1$ at $m=0$, $-2r$, $-4r$, $-6r$, $-8r$, corresponds to the number of massless modes appearing in the Wilson fermion formalism, and the sign of $\delta \nu_\text{4D}$ indeed reflects the chirality of each mode.
This is in contrast to the continuum theory, directly associated with the anomaly, leading to the half-integer topological charge.

\section{Topological phase, domain-wall, and overlap}
\label{sec:top_phase}

\subsection{Domain-wall fermion at the topological insulator boundary}

As discussed in Sec.~\ref{sec:band}, the topology associated with the band structure is well-defined for massive theory, and the topology change occurs at the critical point which is massless $m = 0$.
This means that, according to the band theory, the band topology is only well-defined with an insulator since the mass term plays a role of the band gap.
Nowadays an insulator involving non-trivial topology is called {\it the topological insulator}, in general, and the quantum Hall effect is known as the most fundamental example.

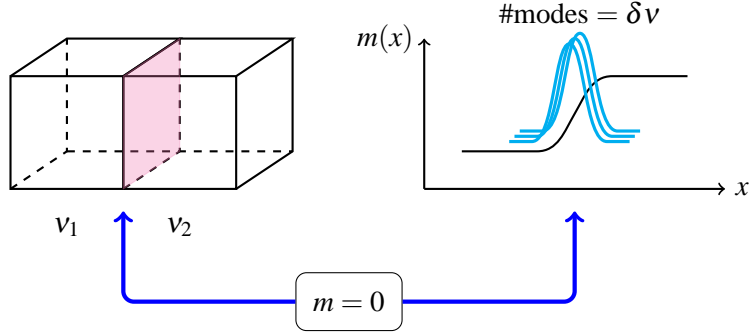
\begin{figure}[t]
  \begin{center}
  \begin{tikzpicture}
   
   \draw [thick] (0,0) rectangle ++(1.5,1.5);
   \draw [thick] (1.5,0) rectangle ++(1.5,1.5);
   \draw [thick] (0,1.5) -- (.75,2) -- (2.25,2) -- (1.5,1.5);
   \draw [thick] (1.5,1.5) -- (2.25,2) -- (3.75,2) -- (3,1.5);
   \draw [thick] (3,0) -- (3.75,.5) -- (3.75,2);
   \draw [thick,dashed] (.75,.5) -- (3.75,.5);
   \draw [thick,dashed] (0,0) -- (.75,.5) -- (.75,2);
   \draw [thick,dashed] (1.5,0) -- (2.25,.5) -- (2.25,2);

   \filldraw[opacity=.3,magenta!70] (1.5,0) -- (2.25,.5) -- (2.25,2) --
   (1.5,1.5) -- cycle;
   
   \node at (.75,-.5) {$\nu_1$};
   \node at (2.25,-.5) {$\nu_2$};

   \begin{scope}[shift={(6,0)}]
    
    \draw [thick,->] (-.5,0) -- (-.5,2) node [left] {$m(x)$};
    \draw [thick,->] (-.5,0) -- (3.5,0) node [right] {$x$};

    \draw [thick] (0,.5) -- (1,.5) to [out=right,in=240] (1.5,1) to
    [out=60,in=left] (2,1.5) -- (3,1.5);

   \draw [very thick,cyan] 
   (.7,.7) -- (1,.7) .. controls (1.3,.7) and (1.3,2) .. (1.5,2) node
   (loc) {} .. controls (1.7,2) and (1.7,.7) .. (2,.7) -- (2.3,.7);

    \begin{scope}[shift={(.07,.07)}]
     \draw [very thick,cyan] 
   (.7,.7) -- (1,.7) .. controls (1.3,.7) and (1.3,2) .. (1.5,2) node
   (loc+) {} .. controls (1.7,2) and (1.7,.7) .. (2,.7) -- (2.3,.7);    
    \end{scope}

    \begin{scope}[shift={(-.07,-.07)}]
     \draw [very thick,cyan] 
   (.7,.7) -- (1,.7) .. controls (1.3,.7) and (1.3,2) .. (1.5,2) node
   (loc-) {} .. controls (1.7,2) and (1.7,.7) .. (2,.7) -- (2.3,.7);    
    \end{scope}    
    
   \end{scope}

   \node [draw, fill=white,
   text width=3em, text centered, rounded corners, minimum height=2em
   ]
   (m0) at (4.5,-1.5) {$m=0$};

   \draw [ultra thick,blue,->,rounded corners] (m0) -- ++(3,0) -- ++ (0,1.2);
   \draw [ultra thick,blue,->,rounded corners] (m0) -- ++(-3,0) -- ++ (0,1.2);

   \draw (loc+) node [above] {\#modes $= \delta \nu$};
   
  \end{tikzpicture}
  \end{center}
 \caption{Domain-wall configuration of the mass term at the topological phase boundary. The localized zero mode on the domain-wall exhibits the gapless edge state of the topological system. The number of edge states is equivalent to the difference of topological charges $\delta \nu = \nu_1 - \nu_2$.}
  \label{fig:DW}
\end{figure}

Typically the topological charge depends only on the sign of the mass term, e.g. \eqref{eq:2D_top} and \eqref{eq:4D_top}, which directly implies that the boundary of the topological insulator, where the topology change must occur, realizes the domain-wall configuration of the mass term.
See Fig.~\ref{fig:DW}.
This can be thought as the origin of the gapless surface (edge) state of the topological insulator.
The number of edge states exactly corresponds to the difference of topological numbers.
See the argument around \eqref{eq:4D_top_lat}.
In this sense, we can compute the topological number in two ways: the characteristic class associated with the bulk system and the number of edge states.
This principle is called the bulk/edge correspondence in the topological phases.
The domain-wall formalism discussed here is a possible way to define a single massless chiral fermion on a lattice, and the existence of odd-number gapless excitations at the topological insulator boundary has been confirmed in several experiments.
See a review article~\cite{Hasan2010} for details.

\subsection{Overlap formula for the topological edge state}

\begin{figure}[t]
 \begin{center}
  \begin{tikzpicture}

   \draw[thick,->] (.3,-1) -- (1.7,-1) node [right] {$p$};
   \draw[thick,->] (-.5,1) -- ++(0,.5) node [left] {$E$};

   \draw[thick] (0,0) to [out=30,in=left] (1,.3) to
   [out=right,in=150] (2,0);

   \filldraw[brown,opacity=.4] 
   (0,0) to [out=30,in=left] (1,.3) to [out=right,in=150] (2,0) --
   (2,-.3) -- (0,-.3) -- cycle;

   \draw[thick] (0,1.5) to [out=-30,in=left] (1,1.2) to
   [out=right,in=210] (2,1.5);

   \filldraw[brown,opacity=.4] (0,1.5) to [out=-30,in=left] (1,1.2) to
   [out=right,in=210] (2,1.5) -- (2,1.8) -- (0,1.8) -- cycle;

   \draw [thick,magenta] (.5,.25) -- (1.5,1.25);
   \draw [thick,magenta] (.5,1.25) -- (1.5,.25);

   \draw[ultra thick,blue,->] (2.7,.75) -- ++ (1,0);

   \begin{scope}[shift={(5,0)}]
    
    \fill
    [top color=red!50!magenta,bottom color=red!10,middle color=red,shading=axis,opacity=0.25]
    (0,0) circle (.5 and 0.1);
    
    \fill
    [left color=red!50!magenta,right color=red!50!magenta,middle color=red!50,
    shading=axis, opacity=0.25] (.5,0) -- (0,.75) -- (-.5,0) arc (180:360:.5 and 0.1);

    \fill
    [top color=red!50!magenta,bottom color=red!50,middle color=red,shading=axis,opacity=0.25]
    (0,1.5) circle (.5 and 0.1);
    
    \fill
    [left color=red!50!magenta,right color=red!50!magenta,middle color=red!50,
    shading=axis, opacity=0.25] (.5,1.5) -- (0,0.75) -- (-.5,1.5) arc (180:0:.5 and 0.1);
    
   \end{scope}

  \end{tikzpicture}
 \end{center}
 \caption{Band spectrum of the topological system. There is an edge state interpolating the bulk valence and conduction bands. The effective theory for the massless surface state is obtained by subtracting massive bulk spectra. The resulting formula is the overlap fermion.}
 \label{fig:edge}
\end{figure}
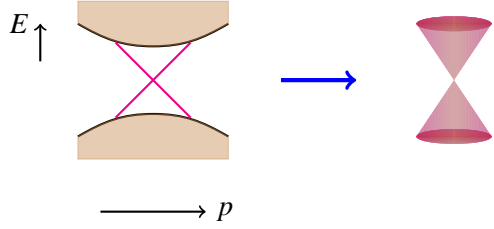

There are various interesting phenomena for the topological insulator surface state from both theoretical and experimental point of view: the Tomonaga--Luttinger liquid behavior of the quantum Hall edge current, the quantum Hall effect at the three-dimensional topological insulator surface, and so on.
In order to study such phenomena peculiar to the topological insulator boundary, it is required to construct an effective theory of the topological surface state. 
Fig.~\ref{fig:edge} shows a band spectrum of the topological system with the open boundary condition so that the non-trivial edge state exists at the boundary.
This is a typical situation for the domain-wall fermion.
In addition to the gapless mode corresponding to the edge state, there are massive bulk spectra to be removed to obtain the edge effective theory.
The prescription to obtain such an effective theory is nothing but the overlap formula~\cite{Neuberger1998}:
\begin{align}
 \det D_\text{eff} =
 \frac{\det D_\text{open}}{\det D_\text{period}}
 \, .
\end{align}
The meaning of this formula is as follows:
The Dirac operator $D_\text{open}$ exhibits the spectrum with the gapless edge state due to the open boundary condition, as depicted in Fig.~\ref{fig:edge}.
The operator in the denominator $D_\text{period}$ comes from the auxiliary (bosonic) degrees of freedom to remove the massive modes.
The latter one is introduced with the periodic boundary condition, and thus it has no edge excitation.
Since the boundary condition does not affect the bulk spectra except for the edge state, the remaining degrees of freedom in the effective operator $D_\text{eff}$ describes the gapless edge state.
In this sense the topological edge state is a physical realization of the overlap fermion.

Once we identify the topological edge state as the overlap fermion, we can apply its several interesting properties to the topological insulator.
The first example is the Ginsparg--Wilson (GW) relation~\cite{Ginsparg1982}:
\begin{align}
 \gamma_5 D + D \gamma_5 = a \, D \gamma_5 D
 \, .
 \label{eq:GW_rel}
\end{align}
A direct consequence of this GW relation is asymmetric chiral transformations of $\psi$ and $\bar{\psi}$~\cite{Luscher1998},
\begin{align}
 \psi \ \to \ e^{i \theta \Gamma_5} \psi \, ,
 \qquad
 \bar{\psi} \ \to \ \bar{\psi} e^{i \theta \gamma_5} \, ,
 \qquad
 \Gamma_5 = \gamma_5 \left( 1 - a D \right)
 \, .
\end{align}
This implies the particle-antiparticle symmetry is broken in the lattice spacing order $O(a)$.
This is a natural realization of the chiral anomaly on a lattice.

In order to generalize this argument to other dimensions, we introduce another expression of the GW relation without using the chiral operator $\gamma_5$,
\begin{align}
 D + D^\dag = a \, D^\dag D
 \label{eq:GW_rel2} 
\end{align}
where we just assume the $\gamma_5$-Hermiticity, $\gamma_5 D \gamma_5 = D^\dag$.
Since this expression does not use the chiral operator $\gamma_5$ explicitly, it is applicable also in odd dimensions.
A solution to this GW relation is obtained with a unitary operator $V$ as follows,
\begin{align}
 D = \frac{1}{a} \left( 1 - V \right) \, .
\end{align}

To apply this relation to the three-dimensional topological insulator and its surface state, let us consider $(2+1)$-dimensional GW relation.
In this case, we obtain a similar asymmetric behavior between $\psi$ and $\bar{\psi}$ under the parity transformation~\cite{Bietenholz2001},%
\begin{align}
 \psi \ \to \ i \, R \, V \psi \, , \qquad
 \bar\psi \ \to \ i \bar{\psi} R
 \label{eq:3D_asym}
\end{align}
and vice versa.
Here we introduced the reflection operator $R$: $(x,y,z) \to (-x,-y,-z)$ with
${R} \, D \, {R} = D^\dag$.
This is a realization of the parity anomaly on a lattice, which is similar to the chiral anomaly discussed above.
In this case we observe the asymmetry under the parity transformation, and in this way, we can apply this argument to arbitrary topological phases associated with the quantum anomaly with the corresponding symmetry and dimensions.

Let us comment on observation of such an asymmetric behavior in experiments.
In the three-dimensional topological insulator surface, the quantum Hall effect can be induced by doping the magnetic impurity, which breaks the time-reversal symmetry.
In this case, if we have the particle-antiparticle (hole) asymmetry, anomalous peak shift of the magneto-optical conductivity $\real \sigma_{xx}(\omega)$ will be observed, which is peculiar to the zero energy state~\cite{Tabert2015}.
While the authors of Ref.~\cite{Tabert2015} explicitly introduce the mass term, which leads to the asymmetry, our argument based on the GW relation could provide more systematic explanation to such an anomalous behavior, for example, as a result of the discretization error.

\section{Topological phases with additional symmetry}
\label{sec:top_add_symm}

In addition to the topological phases, based on the well-known periodic table with respect to the dimension and symmetry~\cite{Schnyder2008,Kitaev2009}, recently several topological systems associated with additional symmetry, e.g. symmetry of the lattice, gain a great interest in condensed-matter physics.
In this section, we demonstrate that the argument discussed in Sec.~\ref{sec:top_phase} is also applicable even in the topological phase with the additional symmetry.

Let us consider the three-dimensional system with the reflection symmetry in $x$-direction.
We define the $x$-reflection operator ${R}_x$: $(x,y,z) \to (-x,y,z)$.
In this case, the invariant plane $x = 0$ under this reflection plays an important role:
The Dirac operator has a symmetry with this operation,
\begin{align}
 {R}_x \, D(x=0,y,z) \, {R}_x = D(x=0,y,z)
 \, .
\end{align}
Then, defining the three-dimensional chiral operator with the $x$-reflection, $\displaystyle \Gamma_x = i \gamma_x \, {R}_x$,
we obtain the GW relation with respect to this operator,
\begin{align}
 \Gamma_x \, D + D \, \Gamma_x = a \, D \Gamma_x D
 \, .
\end{align}
Since we now have the chiral operator $\Gamma_x$, the topological classification is modified as $\mathbb{Z}_2$ (parity) $\to$ $\mathbb{Z}$ (chiral).
This is known as the topological crystalline insulator~\cite{Ando2015}.
Naively speaking, the origin of this modification is understood as follows:
Due to the additional reflection symmetry, a sort of dimensional reduction occurs at the invariant plane with respect to the reflection symmetry.
In this case, therefore, the original three-dimensional system is reduced to the two-dimensional one at the $x=0$ plane, and the operator $\Gamma_x$ plays a role of the two-dimensional chiral operator.

\subsection*{Acknowledgements}

TK would like to thank T. Morimoto for collaboration at early stages of this work.
The~work~of~TK was supported in part by JSPS Grant-in-Aid for Scientific Research (No.~13J04302)~from~MEXT~of~Japan.


\end{document}